\documentclass[12pt,preprint]{aastex}

\shorttitle{Deep near-infrared observations of W3 Main star forming region}
\shortauthors{Ojha et al.}

\begin{document}

\title{Deep near-infrared observations of W3 Main star forming region}

\author{D.K. Ojha\altaffilmark{1}, M. Tamura, Y. Nakajima, and M. Fukagawa}
\affil{National Astronomical Observatory of Japan, Mitaka, Tokyo 181-8588, Japan}
\email{ojha@optik.mtk.nao.ac.jp}
\and
\author{K. Sugitani}
\affil{Institute of Natural Sciences, Nagoya City University, Mizuho-ku, 
Nagoya 467-8501, Japan}
\and
\author{C. Nagashima, T. Nagayama, T. Nagata, and S. Sato}
\affil{Department of Astrophysics, Faculty of Sciences, Nagoya University,
Chikusa, Nagoya 464-8602, Japan}
\and
\author{A.J. Pickles}
\affil{Institute for Astronomy, University of Hawaii, Hilo, HI 96720, USA}
\and 
\author{K. Ogura}
\affil{Kokugakuin University, Higashi, Shibuya-ku, Tokyo 150-8440, Japan}

\altaffiltext{1}{On leave from Tata Institute of Fundamental Research, 
Mumbai (Bombay) - 400 005, India}

\begin{abstract}

We present a deep JHK$_s$-band imaging survey of the W3 Main star forming
region, using the near-infrared camera, SIRIUS (Simultaneous three-color 
InfraRed Imager for Unbiased Surveys), mounted on the University of Hawaii 
2.2 m telescope. The near-infrared survey covers an area of 
$\sim$ 24 arcmin$^2$ with 10 $\sigma$ limiting magnitudes of $\sim$ 19.0, 
18.1, and 17.3 in  J, H, and K$_s$-band, respectively. We construct JHK 
color-color and J/J-H and K/H-K color-magnitude diagrams to identify 
young stellar objects and estimate their masses. Based on these color-color 
and color-magnitude diagrams, a rich population of YSOs is identified which 
is associated with the W3 Main region. A large number of previously 
unreported red sources (H-K $>$ 2) have also been detected around W3 Main. We 
argue that these red stars are most probably pre-main sequence stars with 
intrinsic color excesses. We find that the slope of the K$_s$-band luminosity 
function of W3 Main is lower than the typical values reported for the young 
embedded clusters. 
The derived slope of the KLF is the same as that
found by Megeath et al. (1996), from which analysis by Megeath et al.
indicates that the W3 Main region has an age in the range of 0.3--1 Myr.  
Based on 
the comparison between models of pre-main sequence stars with the observed 
color-magnitude diagram we find that the stellar population in W3 Main is 
primarily composed of low mass pre-main sequence stars. We also report the 
detection of isolated young stars with large infrared excesses which are 
most probably in their earliest evolutionary phases.

\end{abstract}


\section{Introduction} 

The W3 giant molecular cloud (GMC) complex is located in the Perseus spiral 
arm at a distance of 1.83$\pm$0.14 kpc from the Sun (Imai et al. 2000). 
W3 GMC hosts two massive and active star forming regions, W3 Main in the 
north, and W3 (OH) in the south. The W3 Main star forming region contains 
objects such as H II regions, embedded infrared sources (including the 
extremely luminous cluster of sources W3 IRS 5), OH and water masers 
(Wynn-Williams et al. 1974, Forster et al. 1977, Gaume \& Mutel 1987),
which are in different stages of evolution.
 
The millimeter continuum observations have shown the existence of two dense
clumps of about 2000 M$_{\odot}$ associated with the luminous infrared
sources IRS 4 and IRS 5 in the W3 core (Richardson et al. 1989).
VLA observations have resolved the IRS 5 region into a cluster of seven
distinct centimeter radio sources (Tieftrunk et al. 1997, hereafter TGC97;
Claussen et al. 1994, hereafter CG94).   
TGC97 proposed that the spatial and kinematic relation of 
the compact, ultracompact, and hypercompact radio continuum regions toward W3 
Main is
indicative of sequentially triggered star formation caused by the pressure
of the expanding H II regions and the subsequent compression of the molecular
gas. Recently, high resolution continuum imaging at 1.3 and 0.7 cm of four
hypercompact H II regions in W3 IRS 5, suggested that these sources
contain B or O stars (Wilson et al. 2003). 

From a recent near-infrared (NIR) survey of a 
$\sim$ 1\arcmin.5$\times$1\arcmin.5 region towards W3 Main, 
Megeath et al. (1996) found a dense concentration of stars in the molecular
clump surrounding W3 IRS 5. These data showed a large, embedded population of
intermediate to low mass stars co-existing with recently formed OB stars. 
They also argued that the formation of high mass stars is associated with the 
formation of dense clusters of low mass stars in the W3 Main star forming 
region.
Several X-ray sources were detected in the W3 core by Chandra X-ray Observatory
(Hofner et al. 2002). Most of these sources are located at the peak radio 
positions of the W3 H II regions. Hofner et al. (2002) postulated that 
the X-ray sources are the young massive stars that are also responsible for 
the ionization of the compact and ultracompact H II regions in the W3 core.
 
In this paper we present deep J, H, and K$_s$-bands NIR 
observations of the W3 Main star forming region. In comparison with the 
previous NIR survey (Megeath et al. 1996), our survey covers a larger
area ($\sim$ 24 arcmin$^2$) surrounding W3 IRS 5, including compact 
H II regions W3 A, W3 B and W3 D, diffuse H II regions W3 H, W3 J and W3 K, 
and the cometary ultracompact
(UC) H II regions W3 C, W3 E and W3 F. These individual H II regions have 
been labeled following the scheme introduced by Wynn-Williams (1971) and
Harris \& Wynn-Williams (1976). Our motivation is to look for  
new young stellar objects (YSOs) associated with the W3 Main region, 
to determine their evolutionary stages, and to discuss their nature.
Tieftrunk et al. (1998) have presented the three 10\arcmin$\times$10\arcmin 
~mosaics ($\sim$ 300 arcmin$^2$) in K' filter of W3 region. This mosaic has 
similar spatial resolution and depth to the SIRIUS K$_s$ image and covers the 
region stretching from W3 Main to W3(OH). However, due to the crowded nature 
of the sources and the large pixel size, the K'-band mosaic was not suitable 
for photometry of the embedded stellar clusters. 
In Sects. 2 and 3 we
present the details of observations and data reduction procedures, 
Sect. 4 deals 
with the results and discussion and we summarize our conclusions in Sect. 5. 

\section{Observations}

The deep imaging observations of the W3 Main star forming region in the NIR
wavelengths J ($\lambda$ = 1.25 $\mu$m), H ($\lambda$ = 1.65 $\mu$m),
and K$_s$ ($\lambda$ = 2.15 $\mu$m) were obtained on 2000 August 18 
with the University of Hawaii 2.2 m telescope and SIRIUS (Simultaneous 
three-color InfraRed Imager for Unbiased Surveys), a three-color simultaneous 
camera equipped with
three 1024$\times$1024 HgCdTe arrays. The field of view in each band is
$\sim$ 4\arcmin.9 $\times$ 4\arcmin.9, with a pixel scale of 0\arcsec.28
at the Cassegrain focus of $f/10$. The HgCdTe arrays work linearly within
3\% upto 15,000 ADU and saturate at $\sim$ 25,000 ADU 
(Nagayama et al. 2003). At our K$_s$ = 12 mag, the ADU counts are well 
below 15,000, thus we consider the source magnitudes to be correct 
within 3\%.
Further details of the instrument are given in Nagashima et al. (1999) and
Nagayama et al. (2003). 

We obtained 28 dithered exposures of the target centered at 
($\alpha$, $\delta$)$_{2000}$ = ($02^h25^m37^s.00$, 
+62$^{\circ}05^{\arcmin}50^{\arcsec}.0$), 
each 30s long, simultaneously for each band and 18 dithered sky frames
centered at 
($\alpha$, $\delta$)$_{2000}$ = ($02^h25^m22^s.00$, 
+62$^{\circ}34^{\arcmin}41^{\arcsec}.0$),
which is $\sim$ 30\arcmin~north of the target position. The sky frame 
was also used as a reference field for
W3 Main to assess the stellar populations within the W3 Main star forming 
regions (see Sect. 4). Total on-target
integration time in each of the bands was 14 minutes. All the observations 
were done
under good photometric sky conditions. We found an rms magnitude fluctuation 
of less
than 0.05 mag in K$_s$-band during $\sim$ 1 hour of the observations. 
The average seeing in 
the K$_s$-band was 1\arcsec.2 during the observations. The observations were
made at air masses between 1.5 and 1.8. Dark and dome flats were
obtained at the beginning and end of the observations. The photometric
calibration was obtained by observing the standard star 9183 
in the faint NIR standard star catalog of Persson et al. (1998)       
at air masses closest to the target observations.

\section{Data Reduction}

Data reduction was done using the pipeline software based on NOAO's 
IRAF\footnote{IRAF is distributed by the
National Optical Astronomy Observatory, which is operated by the
Association of Universities for Research in Astronomy, Inc., under contract
to the National Science Foundation.} package tasks. Dome flat-fielding and
sky subtraction with a median sky frame were applied. Identification and
photometry of point sources were performed by using the DAOFIND and DAOPHOT
packages in IRAF, respectively. Because of source confusion and nebulosity
within the region, photometry was obtained using the point-spread function 
(PSF) algorithm ALLSTAR in the DAOPHOT package (Stetson 1987). For the 
JHK$_s$-band images the adopted fitting radii were 5 pixels
($\sim$ 1 FWHM of the PSF), and the PSF radius was 21 pixels. The local sky was
evaluated in an annulus with an inner radius of 20 pixels and a width of
35 pixels. We used an aperture radius of 5 pixels ($\sim$ 1\arcsec.4) with 
appropriate aperture corrections per band for the final photometry.


The resulting photometric data are in the
SIRIUS system. For the purposes of plotting these data in 
color-color and color-magnitude diagrams,
we have converted them into the California Institute of Technology (CIT) system
using the color transformations between the SIRIUS and CIT systems
(Nagashima et al. 2003)\footnote{also available at 
http://www.z.phys.nagoya-u.ac.jp/sirius/about/color.html}, which 
have been obtained by observing several of the red standard stars of
Persson et al. (1998). Absolute position calibration was achieved using the    
coordinates of a number of stars from the 2MASS catalogue. The position
accuracy is better than $\pm$0.\arcsec7 rms in the W3 Main field, as compared 
to that of the reference field ($\pm$0.\arcsec2). The relatively poorer
positional accuracy in W3 Main is probably due to a spatially 
varying nebulosity in the field.  
 
The completeness limits of the images were evaluated by adding artificial
stars of different magnitudes to the images and determining the fraction
of stars recovered in each magnitude bin. The recovery rate was greater
than 90\% for magnitudes brighter than 17, 16, and 15 in the J, H, and 
K$_s$-bands respectively. The observations are complete (100\%) to the level
of 15, 15 and 14 magnitudes in J, H and K$_s$-bands respectively. The
limiting magnitudes (at 10 $\sigma$) are roughly estimated to be 
$\sim$ 19.0, 18.1, and 17.3 at J, H, and K$_s$-bands, respectively. 
We found that
within the 10 $\sigma$ completeness limit, the accuracy on magnitudes 
of $\sim$ 98\% stars in our sample is better than 0.1 mag. 
The sources are saturated at K$_s$ $<$ 12. For such bright sources, 2MASS
PSC data were used.

We estimated the errors in photometry due to source confusion and
nebulosity through artificial star experiments. 
The difference between the
magnitudes of the added and recovered stars should reflect the effect of
confusion with other stars and nebulosity. We find that for J = 19.0,
H = 18.1, and K$_s$ = 17.3 stars (at our 10 $\sigma$ mag detection limit), the 
rms error of the difference is 0.22, 0.22, and 0.23 mag, respectively.
The rms error of the difference is 0.11, 0.13, and 0.14 mag, respectively 
for J = 17, H = 16, and K$_s$ = 15 stars (corresponding to the 90\% 
completeness level). The error increases rapidly with increasing magnitude 
(see Fig. 6).

\section{Results and Discussion}

\subsection{Morphology}

The J, H, and K$_s$-band images of the W3 Main star forming region are shown 
in Fig. 1. The circle of a radius of 30\arcsec~(0.27 pc) marked in the H-band 
image represents the cluster region 
around W3 IRS 5 source which has been detected only in the K$_s$-band 
(see Sect. 4.6.1).
The individual compact \mbox{H II}  regions, UC H II regions, and embedded 
IR sources are marked in the K$_s$-band image. 
The K$_s$-band image shows the highest density of stars around IRS 5 molecular 
clump, whereas the density enhancement vanishes in the J-band image. 

Fig. 2 shows a composite JHK$_s$ color image (J represented in blue, H in 
green, and K$_s$ in red) of the W3 Main star forming region.  
The image shows bright nebulosities towards the compact H II regions W3 A, 
W3 B, and W3 D. We also detect a faint nebulosity around the UC H II regions
W3 C, W3 E, W3 F, and W3 G. W3 H, W3 J, and W3 K are more diffuse 
and dispersed H II regions. The bright blue massive stars of
spectral types O and B (TGC97) are the ionizing sources
of the compact and diffuse H II regions.  
Dark filaments extending from north-west to south-east can be seen between
the diffuse nebulosity found throughout the whole image.
Note the very red object near the center of the image, \mbox{W3 IRS 5}
($\alpha_{2000}$ = 2$^h$25$^m$40$^s$.76, 
$\delta_{2000}$ = +62$^{\circ}$05\arcmin52\arcsec.5), a deeply
embedded infrared source which has been detected only in K$_s$-band.
We see a dense cluster of embedded stars surrounding
IRS 5. A large number of red young stars are 
also seen around \mbox{IRS 5}, which are presumably embedded in the
molecular core (see Sect. 4.3).  

\subsection{Photometric analysis}

\subsubsection{Color-Color diagram}

We obtained photometric data of 986 sources in J, 1236 in H, and 1512 in
K$_s$-band. The drastic drop in the number of sources detected at the shorter
wavelengths in spite of greater sensitivity gives the first indication of the
extremely high interstellar extinction around the W3 Main region.   
Figs. 3a and b show the J-H/H-K color-color (CC) diagrams of the W3 Main
star forming region and the reference field, respectively, for the sources 
detected
in the JHK$_s$ bands with a positional agreement of less than 3\arcsec
~and with photometric errors in each color of less than 0.1 mag. 
The reference field is also used for the correction of field star 
contamination 
from the raw K$_s$-band luminosity function of W3 Main (see Sect. 4.4). 
In Figs. 3a and b, the solid
and broken heavy curves represent the unreddened main sequence and
giant branches (Bessell \& Brett 1988) and the parallel dashed lines are the
reddening vectors for early and late type stars (drawn from the base and tip
of the two branches). The dotted lines indicate the locus of T-Tauri stars
(Meyer et al. 1997). We have assumed that A$_J$/A$_V$ = 0.282,
A$_H$/A$_V$ = 0.175, and A$_K$/A$_V$ = 0.112 (Rieke \& Lebofsky 1985).
As can be seen in Fig. 3a, the stars in W3 Main are distributed
in a much wider range than those in the reference field (Fig. 3b),
which indicates that a large fraction of the observed sources in W3 Main
exhibit NIR excess emission. 
We classified the sources into three regions in the CC diagram 
(see e.g. Tamura et al. 1998, Sugitani et al. 2002). ``F'' sources
are located between reddening vectors projected from the intrinsic color
of main-sequence stars and giants and are considered to be unreddened and 
reddened field stars
(main-sequence stars, giants), or Class III / Class II sources having small
near-infrared excess.
``T'' sources are located redward of region F but blueward of the reddening
line projected from the truncated point of the T Tauri locus of 
Meyer et al. (1997). These sources are considered to be classical T-Tauri
stars (Class II objects). 
``P'' sources
are those located in the region redward of region T and are most likely
Class I objects. In Fig. 3a, a gap is seen between the reddened stars and
unreddened stars (located near the main-sequence locus : H-K $\sim$ 0.4, 
J-H $\sim$ 1.0).
By dereddening the stars on the CC diagram that fell within the reddening
vectors emcompassing the main sequence and giant stars, we found the 
amount of visual extinction (A$_V$) for each star. The individual extinction 
values range from 0 to 24 magnitudes with an average extinction of
A$_V$ $\sim$ 8 mag. The stars lying on the lower left side of the reddening 
band are mostly foreground stars as supported by their low values of A$_V$.

\subsubsection{Color-Magnitude diagram}

The color-magnitude (CM) diagram is a useful tool to study the nature of the
stellar population within star forming regions and also to estimate
its spectral types. In Fig. 4, the H-K vs K CM diagram for all
the sources detected in JHK$_s$ bands, plus some 384 stars fainter than our
limit at J-band but still above the detection threshold in H and K$_s$-bands
are plotted. The vertical solid lines (from left to right in 
Fig. 4) represent the main sequence curve reddened by 
A$_V$ = 0, 20, 40 and 60 magnitudes,
respectively. We have assumed a distance of 1.83 kpc to the sources to reproduce
the main sequence data on this plot. The parallel slanting lines in Fig. 4
trace the reddening zones for each spectral type. YSOs (Class II and I) 
found from the CC
diagram (Fig. 3a) are shown as stars and filled triangles. However, it is 
important to note that even those stars not shown with stars or filled
triangles may also be YSOs with an intrinsic color excess, since those stars
are detected in the H and K$_s$-bands only 
and are not in the J band due to their very red
colors. Two bright and very red objects (K $<$ 11.5, H-K $>$ 2.8) located
in the upper right corner of the figure are probably the very young stars 
in their earliest evolutionary phases (see Appendix and Fig. 10c and 10d). 
The bright infrared sources labeled with IRS numbers (Fig. 4) are 
associated with the molecular clumps and H II regions. These sources are 
shown in Table 1.  

\subsection{Spatial distribution of YSOs and cool red sources}

In our deep NIR observations $\sim$ 40 very red sources are detected only in 
the H and K$_s$-bands. These sources have colors redder than H-K $>$ 2
in Fig. 4. They are also YSO candidates. 
In Fig. 5, the spatial distribution of YSO candidate sources identified
in Figs. 3a and 4 are shown. Stars represent 
sources of T-Tauri type (Class II), filled 
triangles indicate Class I sources, and filled circles denote 
the very red sources (H-K $>$ 2).
 
In general, Class I and Class II candidates 
(located in the T and P regions in Fig. 3a) are distributed all over the 
field, 
however there is an apparent concentration of these sources around W3 A, 
W3 B, W3 F, W3 H, and W3 J H II regions as seen in Fig. 5. 
What is particularly striking is that most of these YSOs are 
associated with the diffuse ionized gas at the edge of the compact 
H II regions.   
Stars with large color indices (H-K $>$ 2) are seen near the dense parts of 
the molecular cloud. Most of them are clustered near the 
massive molecular clumps surrounding the luminous infrared sources W3 IRS 4
and W3 IRS 5. Some of them are expected to be members of the embedded stellar
cluster around W3 IRS 5. It is to be noted that these red sources are not
associated with any H II regions (except the two sources south of W3 A).
Therefore, these sources associated with the molecular clumps are 
embedded PMS stars, presumably.    


The average extinction through the molecular cloud around 
IRS 5 that hosts the embedded cluster is A$_V$ $\sim$ 15 mag. If we assume 
that the large H-K ($>$ 2) color results merely from interstellar 
reddening affecting normal stars,  
then the extinction value might even exceed 40 mag in the molecular 
cloud where most of the red
stars are found. However, with such a large amount of absorption, diffuse
emissions are unlikely to be detected in the NIR. Since most of the red sources
are associated with faint diffuse emission, this provides an evidence that 
these sources are YSOs with intrinsic NIR excess and possibly 
local extinction also. 
In Fig. 4, a large fraction ($\sim$ 94\%) of these sources are located above 
the straight 
line drawn from an A0 star parallel to the extinction vector.
This suggests that they are high mass stars with circumstellar 
materials.


\subsection{The K$_s$-band Luminosity Function}

We use the K$_s$-band luminosity function (KLF) to constrain the
initial mass function (IMF) and age of the embedded stellar population in 
W3 Main. 
To derive the KLF, we have determined the completeness of the data 
through artificial star experiments using {\it addstar} in IRAF. This was
performed by adding fake stars in random positions into the images at 
0.5 magnitude intervals and then checking how many of the added stars could be 
recovered at various magnitude intervals. We repeated this procedure 
at least 8 times. We thus obtained the detection rate as a function of 
magnitude, 
which is defined as the ratio of the number of recovered artificial stars 
over the number of added stars. Fig. 6 shows the rms error in the magnitude
(difference between the magnitudes of the added and recovered stars)
of the recovered fake stars as a function of the recovered magnitude. 
At K$_s$ = 17.3 (10 $\sigma$ detection limit), the rms error is 0.23 mag
(see Sect. 3).
The error is larger for the cluster than for the whole region due to source
confusion and nebulosity in the cluster region. In Fig. 7, we present the raw KLFs for the
cluster of a radius of 30\arcsec~(0.27 pc) around W3 IRS 5 (see Fig. 1)
and the whole W3 Main region along with the photometric
completeness determined as above. We find that the completeness 
ratio drops below 90\% at 14th magnitude 
and is lower for the cluster than for the whole region.

In order to estimate the foreground and background contaminations, we made use 
of both the galactic model by Wainscoat et al. (1992) and
the reference field star count. The star counts were 
predicted in the direction of the reference field close to the W3 Main 
(see Sect. 2), which is also corrected for the photometric completeness. 
From Fig. 4, we conclude that the average
extinction to the embedded cluster is A$_V$ $\sim$ 15 (H-K = 1). Assuming
spherical geometry, then background stars are seen through A$_V$ $\sim$ 30.
Therefore, in simulating the background of the region, we added an extinction 
value of A$_V$ = 30 mag (or A$_K$ = 3.36 mag) to the background stars. 
Then we scaled the model prediction to the star counts in the reference 
field, and subtracted the combined foreground (d $<$ 1.8 kpc) and 
background (d $>$ 1.8 kpc with A$_K$ = 3.36 mag) data from the KLF of the 
W3 Main region.


After correcting for the foreground and background star contamination and 
photometric 
completeness, the resulting KLFs are presented in Fig. 8 for the cluster and 
whole W3 Main regions. Both the KLFs follow power-laws in shape. We estimate 
a total of 156 sources with K $<$ 17.5 in the cluster region after applying 
the corrections for completeness and foreground and background star 
contamination.
Given a distance of 1.83 kpc and assuming
that the cluster is spherical with an apparent radius of 30\arcsec~(0.27 pc), 
we derive a cluster density of 
$\sim$ 2000 stars pc$^{-3}$ for K $<$ 17.5.
The observed density of other embedded clusters is typically lower than
$\sim$ 1000 stars pc$^{-3}$ (Lada \& Lada 2003; Carpenter et al. 1993). 
The high 
density of stars in W3 Main may be a result of the youth of the cluster
and the enormous mass of molecular gas available in the W3 Main molecular
core (Megeath et al. 1996). 

In Fig. 8, a power-law with a slope $\alpha$ 
($d N(m_K)/dm_K \propto 10^{\alpha m_K}$, where $N(m_K)$ is the number of
stars brighter than $m_K$)
has been fitted to each KLF using a linear least-squares fitting routine. 
The derived power-law slopes for various regions in W3 Main are shown in
Table 2. We find that our estimate of the power-law slopes is in remarkable 
agreement with that of Megeath et al. (1996) in spite of the larger 
survey area which includes objects
such as compact H II, UC H II, and diffuse H II regions. Our results therefore 
confirm that the KLF of the W3 Main region shows a power-law slope which 
is lower than those generally reported for the young embedded 
clusters ($\alpha \sim 0.4$, e.g. Lada et al. 1991, 1993; Lada \& Lada 2003). 
Thus, this low value of the slope is
indeed an intrinsic property of the stellar population in this region.
As shown in detail by Megeath et al. (1996), the estimated KLF slopes of 
the whole W3 Main region are roughly consistent with the Miller-Scalo IMF
if the age of W3 population is $\sim$ 0.3--1 Myr.

\subsection{Mass Estimation}

Fig. 9 shows the CM diagram (J-H vs J) for $\sim$ 160 YSO candidate sources 
identified in Figs. 3a and 4. We estimate the mass of the sources by comparing 
them with the evolutionary models of PMS stars (Palla \& Stahler 1999). The 
solid curve in Fig. 9 denotes the loci of 10$^{6}$ yr old PMS stars and the 
dotted curve for those of 0.3$\times10^{6}$ yr old ones. Masses range from 
0.1 to 4 M$_{\odot}$ from bottom to top, for both curves. Solid oblique 
reddening line denotes position of PMS with 2 M$_{\odot}$ for 1 Myr and the 
dotted oblique lines denote positions of PMS with 2 and 4 M$_{\odot}$ for 
0.3 Myr, respectively. Most of the objects well above the PMS tracks 
are luminous and massive ZAMS stars (see Table 1 \& Fig. 4).
We use the J luminosity rather than that of H or K$_s$, as J-band is less 
affected by the emission from circumstellar materials 
(Bertout, Basri, \& Bouvier 1988). 

If we assume that the age of the stars in the W3 Main star forming region is 
$\sim$ 0.3 Myr, 86\% of the YSO candidates detected in J, H, and K$_s$-bands 
have masses less than 4 M$_{\odot}$ (Fig. 9)  
and at least 80\% of the stars have masses less than 2 M$_{\odot}$. Even if 
the age of the stars is 1 Myr, 75\% of the stars have masses below 
2 M$_{\odot}$. At the distance of 1.83 kpc, assuming an age in the range 
0.3--1 Myr, and an extinction at K band between 0 and 1 mag 
(up to A$_V$ $\sim$ 10), the magnitude limit (corresponding to the 90\% 
completeness level) corresponds to M $\sim$ 0.4 M$_{\odot}$, 
according to the PMS 
evolutionary tracks from Palla \& Stahler (1999). However, at our 
10 $\sigma$ mag detection limit, the mass would then go down to 
$\sim$ 0.1 M$_{\odot}$. This gives an estimate of the lowest mass limits of 
the detected stars in the W3 Main star forming region in our sample. 

Therefore, the stellar population in W3 Main is primarily composed of low mass 
PMS stars. We also see the presence of lower mass 
stars forming a well defined cluster (e.g. near IRS 5) together with O-B type 
stars which have recently formed. These results support the hypothesis that 
the formation
of high mass stars is associated with the formation of dense clusters of low
mass stars (e.g. Lada \& Lada 1991, Persi et al. 1994, Tapia et al. 1997). 

\subsection{Comments on individual sources}

\subsubsection{Embedded massive YSOs : W3 IRS 5 \& IRS 4}

The infrared sources, W3 IRS 5 
($\alpha_{2000}$ = 2$^h$25$^m$40$^s$.76,
$\delta_{2000}$ = +62$^{\circ}$05\arcmin52\arcsec.5)
and \mbox{W3 IRS 4} 
($\alpha_{2000}$ = 2$^h$25$^m$30$^s$.97,
$\delta_{2000}$ = +62$^{\circ}$06\arcmin20\arcsec.9),
are associated with dense molecular cores. 
The stellar cluster near IRS 5 has a total luminosity of
about 2$\times$10$^5$ L$_{\odot}$. The cluster of hypercompact
continuum sources W3 Ma-g toward IRS 5, with diameters of $<$ 700 AU, 
is situated between W3 A and B
(CG94, TGC97). As seen in Sect. 4.4, we estimate a stellar density
of the cluster around IRS 5 of $\sim$ 2000 stars pc$^{-3}$ (for K $<$ 17.5). 
Radio studies indicate a hydrogen 
density of about 10$^6$ cm$^{-3}$ and a column density of about 
10$^{23}$ cm$^{-2}$ around this source (TGC97). The source
is detected only in our K$_s$-band image (K = 12.29 mag).     

IRS 4 lies $\sim$ 80\arcsec~west of IRS 5 and shows a similarly 
high luminosity. It is associated with
the hypercompact continuum source W3 Ca (TGC97).
At 450 and 800 $\mu$m, the region near IRS 4 is partially resolved into two 
separate
maxima, one near IRS 4 and the other about 20\arcsec~south of IRS 4 
(Richardson et al. 1989, Ladd et al. 1993). None of these FIR sources coincide
with prominent H II regions. The source is detected only in our
H and K$_s$-band images (H = 15.88 mag, K = 12.96 mag).  

Both IRS sources show outflow activity. These regions 
must be very early signposts of recent star formation.
Majority of the sources located in the IRS 5 and IRS 4 regions have masses 
more than 2 M$_{\odot}$ (from Fig. 9). They are also located above the
straight line drawn from an A0 star parallel to the extinction vector
(Fig. 4), indicating the OB star nature. 

\subsubsection{The ultracompact H II regions : W3 C, W3 E, W3 F \& W3 G}

The W3 C UC HII region, which almost coincides with IRS 4 at the edge of the 
dense molecular clump, has an irregular brightened edge. W3 C is associated 
with W3 Ca, the hypercompact region north-east of W3 C. Most probably it is 
associated with the 450 $\mu$m sub-mm source SMS 2 (TGC97). Emissions at 5 and 
20 $\mu$m also show a peak at W3 C (Ladd et al. 1993). The morphology of the 
very red nebulosity extending from W3 C / IRS 4 in Fig. 2 suggests that it is
an embedded cometary-shaped H II region. The 6 cm VLA maps also show a
cometary shape. 

We find that a source with spectral type B0 
with a shell of diffuse near-IR emission is embedded in W3 E.
We designated this source as IRS N1 (see Fig.4 and Table 1).
The estimated spectral type of the source matches well with that
by TGC97, who find
that the Lyman continuum photon flux of ionized gas for W3 E indicates
a B0.5--B0 ZAMS.      

W3 F lies south-west of the dense molecular dust cloud centered on IRS 5.
It is coincident with a deeply embedded infrared source IRS 7, 
which is detected only in the H and K$_s$-band images. 
The color and IR luminosity of this source indicate that it is a B0--B1 type 
star (Table 1). From the radio data TGC97 find that a ZAMS star of B0--O9.5 
spectral type is ionizing W3 F.

The W3 G UC H II region is seen in radio between the two dense molecular 
condensations of the W3 core (TGC97). There do not appear to be any NIR sources
that are associated with W3 G; however TGC97 deduced the ionizing source of 
spectral type B0.5--B0 ZAMS from the radio study. They argued that perhaps 
such an ionizing source is located behind the dense molecular gas and 
therefore not detected in IR.   

\subsubsection{The compact H II regions : W3 A, W3 B \& W3 D}

W3 A is a shell-like, asymmetrically edge-brightened H II region
and most likely is in a late stage of its expansion (TGC97). The main
part of W3 A is no longer embedded in the dense molecular gas 
(Roberts et al.  1997). The embedded O5--O6 stars W3 IRS 2 
and IRS 2a 
are located close to the center of W3 A. The O6 star IRS 2b 
lies in the NW of W3 A. The less luminous O9--B0 star IRS 2c 
is located $\sim$ 40\arcsec~east of IRS 2. The spectral types of these objects 
are estimated from the CM diagram (Fig. 4) based on their colors and IR 
luminosities. These massive stars are collectively responsible for the 
ionization of the W3 A H II region. Despite the number of these bright sources,
we find a low number of low-mass YSOs in this H II region (Fig. 5), which may 
be due to confusion with the high surface brightness of the diffuse nebulosity 
of W3 A.

W3 B is an asymmetrically bright shell-like H II region. It is located between
the two dense molecular condensations. A bright compact O6 type source
IRS 3a 
is associated with W3 B, which is 
responsible for ionizing the W3 B H II region. This source is detected only
in our H and K$_s$-band images. 
W3 B is most likely an emerging
blister only partly embedded in the molecular cloud (TGC97).

W3 D is a weak and extended H II region. It is associated with the infrared
source IRS 10 (Dyck \& Simon 1977). This source is clearly detected in 
our NIR images. The measured magnitudes for this source
are 17.45, 14.86, and 12.98 in J, H, and K-bands, respectively. The color
and IR luminosity of this source indicate that it is a B2 type star, which 
is however 
not consistent with the spectral type derived from the radio observations
(TGC97). 

\subsubsection{The diffuse H II regions : W3 H, W3 J \& W3 K}

The diffuse H II region W3 H is located north of the dense molecular gas.
The probable exciting star of W3 H is very bright in all the three bands.
The color and IR luminosity of this source indicate that it is a B0--B1 type
star with a visual extinction of A$_V$ = 6 - 12. 
We call this source as IRS N2 (see Fig. 4 and Table 1).  

W3 J and W3 K are very diffuse and much more dispersed H II regions located
toward the south of the W3 Main core. These regions are older than the core 
regions as lack of dense molecular meterial is seen around them. The ionizing
sources of W3 J with spectral type B1--B2 and of W3 K with spectral type B0
may have already dispersed the molecular material. These sources are named
as IRS N3 and IRS N4, respectively (see Fig. 4 and Table 1). 
A concentration of YSOs
is seen to the south-east of W3 J where edge-brightening is still
observed (TGC97). Most of these YSOs associated with W3 J and W3 K are fainter 
and have masses lower than the YSOs associated with other compact H II 
regions.    

\subsection{Star formation toward W3 Main}

The three adjoining regions in the Perseus spiral arm, W3, W4, and W5 are
all complexes of active star formation identified by H II regions and
aggregates of young stars. Among them the W3 GMC is the youngest. The 
cluster of compact, ultracompact, and hypercompact H II regions embedded 
within the W3 molecular cloud appears to be ionized by a recently formed 
association of O and B stars (TGC97). From our NIR study it appears that 
the W3 Main region contains both B stars and lower mass stars continuously 
forming. 

It is generally agreed that W4 (= IC 1805), located to the east of W3, was 
the first of the three large H II regions to be formed and that its expansion 
might have recently triggered star formation towards the W3 molecular cloud 
(Elmegreen \& Lada 1977, Dickel et al. 1980, Thronson et al. 1985,
Carpenter et al. 2000). The close association of the embedded clusters
with adjacent H II regions also suggests that triggering may
have played an important role in the formation of these clusters
(Lada \& Lada 2003). The more isolated W5 (= IC 1848), which is found to the
east of W4, also shows indications of triggered star formation probably on a 
lower level (Karr \& Martin 2003).

In summary, W3 GMC characterized by H II regions, high mass stars,
embedded IR clusters and dense molecular cores, and therefore represents
an important source for the study of star formation. Compared with,
e.g., the Orion Nebula (O'Dell 2001) and M17 (Jiang et al. 2002) regions,
one of the prominent features of the star formation in W3 Main is the
absence of dominant OB stars.

 
\section{Conclusions}

A deep JHK$_s$-band NIR imaging survey of YSOs associated with the W3 Main 
star forming region is presented. The survey covers a 
4\arcmin.9$\times$4\arcmin.9 area down to a 
limiting magnitude (10 $\sigma$) of J = 19.0, H = 18.1, and K$_s$ = 17.3.
From the analysis of these images we derive the following conclusions : 

1) A cluster of YSOs (Class II and Class I sources derived from their NIR 
colors) has been detected in the W3
Main core and near the compact, ultracompact, and diffuse H II regions.

2) A large number of red stars (H-K $>$ 2) are detected in the molecular 
cloud region, most of them clustered around the molecular clumps associated
with IRS 5 and IRS 4. Some of them are also associated with the diffuse 
emission near the dense molecular clumps. We argue that most of the reddest 
stars are YSOs with circumstellar materials.

3) The KLF of the W3 Main region shows the power-law slope : 
$\alpha$ = 0.26$\pm$0.02, which is lower than the typical values reported
for the embedded young clusters. Our finding also confirms the previous 
results of Megeath et al. (1996) for a smaller region around W3 IRS 5.

4) The observed density of the cluster region around W3 IRS 5 is
$\sim$ 2000 stars pc$^{-3}$ for K $<$ 17.5, which is larger than the
typical values ($\sim$ 1000 stars pc$^{-3}$) reported for other embedded 
clusters.    

5) Using the age of W3 Main in the range of 0.3--1 Myr determined by
Megeath et al. (1996), we find that 
about 80\% of the YSO candidates have an upper mass limit of 4 M$_{\odot}$. 
We estimate that the lowest mass limit of Class II \& Class I candidates 
in our observations is 0.1 M$_{\odot}$.
Therefore, the stellar population in W3 Main is primarily
composed of low mass PMS stars. 

\acknowledgments

It is a pleasure to thank the anonymous referee for a most thorough reading
of this paper and several useful comments \&
suggestions, which greatly improved the scientific content of the paper.
We thank the staff of the UH 2.2 m telescope for supporting the first run of 
SIRIUS. DKO was supported by the Japan Society for the Promotion of Science 
(JSPS) through a fellowship during which most of this work was done. We thank
Francesco Palla for providing us with their PMS grids. MT acknowledges 
support by Grant-in-Aid (12309010) from the Ministry of Education, Culture,
Sports, Science, and Technology.  

This publication makes use of data products from the Two Micron All Sky
Survey, which is a joint project of the University of Massachusetts and the
Infrared Processing and Analysis Center/California Institute of Technology,
funded by the National Aeronautics and Space Administration and the National
Science Foundation. 

\vskip 1cm
\begin{center}
APPENDIX
\end{center}

\appendix{\centerline {\bf \large Selected interesting regions}}

 
In Fig. 10 we present some selected areas of the W3 Main star forming region
in our new NIR images that are of noteworthy interest. 

(a) About 40\arcsec~south-east of W3 A, a circular bracelike diffuse emission 
is seen (Fig. 10a). This is probably illuminated by two bright stars of the 
same luminosity located toward east of this feature. A few YSOs with NIR color 
excess are located at the southern edge on the bright lane. 
The dark area is also visible between the bright lane and W3 A, which is seen 
to the north. 

(b) Fig. 10b is a section in the south-east corner of our image, where we 
detect an H$_{2}$ knot 
reminiscent of Herbig-Haro (HH) object (elongated and bright in K$_s$-band) 
which surrounds at least three YSOs. This object was noted already by 
Tieftrunk et al. (1998). Spectra of this knot/jet and another about
2\arcmin~south-west show H$_2$ (1-0) S(1) emission and are probably HH-like
objects
(Tieftrunk et al. 1998).
This object is barely detected in the J-band as well. A bright source in 
K$_s$-band (K = 12.57, H-K = 1.35) is located at the northern tip of the 
object. It is, however, striking that no other shock-excited H$_2$ objects
are found in such an active star forming region as W3 Main.

(c) Fig. 10c shows an isolated red source at the center of our image, which is
located towards the south-west of IRS 4. The source is clearly resolved into 
double stars (A and B) separated by $\sim$ 4\arcsec~in our NIR images. 
Both these sources are detected 
only in our H and K$_s$-band images. The infrared colors of these sources 
indicate that they are very red objects
(H-K = 3.37 for A, and H-K = 2.49 for B). The source A looks extended in the
image, probably implying that at 1.6--2.2 $\mu$m wavelengths, we are looking 
at the radiation from an embedded young star scattered by a dusty circumstellar
envelope. They are most probably very young stars in their earliest 
evolutionary phases.  

(d) Dark filamentary lanes are seen (Fig. 10d) with irregular shapes which 
break a 
more diffuse nebulosity extending throughout the whole W3 Main region. They 
are associated with the dense molecular gas. An infrared source marked by an
arrow with large
color excesses (J-H = 4.77, H-K = 2.98) is located inside the dark lanes.
This is most probably a young star in its earliest evolutionary phase.   


\clearpage

\begin{table}
\begin{center}
\caption{Bright infrared sources associated with the molecular clumps and H II regions}
\begin{tabular}{ccccccc} 
\\
\tableline
\tableline
Source  & RA (2000)   & DEC (2000)        &  J    &   H    & K     & Sp. Type         \\
        & $hh~mm~ss$  & $dd~mm~ss$        &  mag  &  mag   & mag   & (From CM diagram)\\
\tableline
IRS 4         & 02 25 30.97 & +62 06 20.9 &       & 15.88$\pm$0.04 & 12.96$\pm$0.04 & O9-B0\\
IRS 5         & 02 25 40.76 & +62 05 52.5 &       &       & 12.29$\pm$0.02 & \\ 
IRS N1 (W3 E) & 02 25 35.15 & +62 05 34.8 & 17.52 & 13.94$\pm$0.02 & 11.85$\pm$0.02 & B0\\
IRS 7 (W3 F)  & 02 25 40.48 & +62 05 40.3 &       & 16.33$\pm$0.05 & 13.44$\pm$0.03 & B0-B1\\
IRS 2 (W3 A)  & 02 25 44.43 & +62 06 11.7 & 12.06$\pm$0.03 & 10.04$\pm$0.03 &  8.85$\pm$0.03 & O5-O6\\
IRS 2a (W3 A) & 02 25 43.34 & +62 06 15.4 & 12.68 & 11.54$\pm$0.06 &  9.80 & O5-O6\\
IRS 2b (W3 A) & 02 25 41.74 & +62 06 24.5 & 13.64$\pm$0.06 & 11.36$\pm$0.06 &  9.86$\pm$0.05 & O6\\
IRS 2c (W3 A) & 02 25 47.09 & +62 06 13.1 & 13.49 & 11.45 & 10.19$\pm$0.04 & O9-B0\\ 
IRS 3a (W3 B) & 02 25 37.83 & +62 05 52.1 &       & 15.12$\pm$0.02 & 12.19$\pm$0.02 & O6\\
IRS 10 (W3 D) & 02 25 29.80 & +62 06 31.8 & 17.45$\pm$0.05 & 14.86$\pm$0.02 & 12.98$\pm$0.02 & B2\\
IRS N2 (W3 H) & 02 25 32.60 & +62 06 59.7 & 11.69$\pm$0.02 & 10.49$\pm$0.03 &  9.78$\pm$0.02 & B0-B1\\
IRS N3 (W3 J) & 02 25 27.35 & +62 03 43.4 & 11.22$\pm$0.02 & 10.45$\pm$0.02 &  9.99$\pm$0.02 & B1-B2\\
IRS N4 (W3 K) & 02 25 44.85 & +62 03 41.3 & 10.19$\pm$0.02 &  9.35$\pm$0.02 &  8.90$\pm$0.02 & B0\\
\tableline
\end{tabular}
\end{center}
\end{table}

\clearpage

\begin{table}
\caption{Power-law fits to KLFs in W3 Main}
\begin{tabular}{cc} 
\\
\tableline
\tableline
Region   & $\alpha$ \\
\tableline
Cluster  & 0.17$\pm$0.02 \\
Whole W3 & 0.26$\pm$0.01 \\
Whole W3 - Cluster & 0.28$\pm$0.02 \\
\tableline
\end{tabular}
\end{table}

\clearpage

\begin{figure}
\vspace*{-2cm}
\caption{J, H, and K$_s$-band images of the W3 Main star forming region 
displayed in a logarithmic intensity scale. The circle of a radius 
of 30\arcsec~(0.27 pc) marked in the H-band 
image shows the cluster region around W3 IRS 5. The locations of the individual
H II, UC H II regions, and the embedded IR sources are marked in the K$_s$-band
image. North is up and east is to the left. The abscissa and the ordinate are 
in J2000.0 epoch.
\label{fig1}}
\end{figure}

\clearpage 

\begin{figure}
\caption{JHK$_s$ three-color composite image of the W3 Main star forming region
(J: blue, H: green, K$_s$: red) obtained by SIRIUS mounted
on the UH 2.2 m telescope. The field of view is 
$\sim$ 4\arcmin.9$\times$4\arcmin.9. North is up and east is to the left.
\label{fig2}} 
\end{figure}

\clearpage

\begin{figure}
\caption{Color-color diagrams of (a) the W3 Main star forming region and 
(b) the reference field for the unsaturated sources (K$_s$ $>$ 12) 
detected in JHK$_s$-bands with photometric errors less than 0.1 mag. 
In (a), open squares show the saturated SIRIUS sources (K$_s$ $<$ 12), 
which have been replaced by the corresponding 2MASS sources. Open triangles
indicate 2MASS sources with upper limits in magnitudes. A gap is seen between
the reddened stars and unreddended stars (located near the main-sequence
locus : H-K $\sim$ 0.4, J-H $\sim$ 1.0). The sequences for 
field dwarfs (solid curve) and giants (thick dashed curve) are from 
Bessell \& Brett (1988). The dotted line represents the locus of T-Tauri stars 
(Meyer et al. 1997). Dashed straight lines represent the reddening vectors 
(Rieke \& Lebofsky, 1985). The crosses on the dashed lines are separated by 
A$_V$ = 5 mag. 
\label{fig3}}
\end{figure}

\clearpage

\begin{figure}
\caption{Color-magnitude diagram for the sources detected in H and K$_s$-bands 
with photometric errors less than 0.1 mag. The open squares show the
saturated SIRIUS sources (K$_s$ $<$ 12), which have been replaced by the 2MASS 
data. The open triangles indicate 2MASS sources with upper limits in 
magnitudes. Stars and filled triangles represent the YSOs identified from
the regions T and P in Fig. 3a, respectively.  
The vertical solid lines from left to right indicate the  
main sequence track at 1.83 kpc reddened by A$_V$ = 0, 20, 40, and 60 
magnitudes,
respectively. The intrinsic colors are taken from Koornneef (1983). Slanting
horizontal lines identify the reddening vectors (Rieke \& Lebofsky 1985).
Also shown are the positions of known IRS sources.  
\label{fig4}}
\end{figure}

\clearpage

\begin{figure}
\vspace{-4cm}
\caption{Spatial distribution of the
YSO candidates superposed on the K$_s$-band image with a logarithmic 
intensity scale. Stars represent
T-Tauri sources (Class II), filled triangles indicate  
Class I sources, and filled circles denote the red sources (H-K $>$ 2).
The abscissa and the ordinate are in J2000.0 epoch.
\label{fig5}}
\end{figure}

\clearpage
\begin{figure}
\caption{The rms error in the magnitude of the recovered fake stars for the
cluster and whole W3 regions, respectively.
\label{fig6}}
\end{figure}

\clearpage

\begin{figure}
\caption{(a) and (c) The raw K$_s$-band luminosity functions for the cluster 
and whole W3 regions, respectively. The error bars correspond to 
$\pm \sqrt(N)$, where N is the number of stars in each magnitude bin. 
(b) and (d) The detection completeness as a function of magnitude for the 
cluster and whole W3 regions, respectively. The error bars are shown as
$\pm$1 $\sigma$ of the mean of 8 trials performed in each magnitude bin.
\label{fig7}} 
\end{figure}


\clearpage

\begin{figure}
\caption{(a) and (c) The corrected K$_s$-band luminosity functions for the 
cluster and whole W3 regions, respectively. The dotted lines show the KLFs 
corrected only for the completeness. The dashed lines denote the field star 
counts from the reference field corrected for the completeness and 
modified to reflect the unextincted foreground stars
and background stars reddended by A$_V$= 30 (A$_K$ = 3.36), 
with the help of the galactic model (Wainscoat et al. 1992).
The solid lines correspond to the field star 
subtracted KLFs. (b) and (d) The completeness-corrected and field 
star-subtracted KLFs of the cluster and whole W3 regions,
respectively. The solid lines are the best linear fit to the data points.
\label{fig8}}
\end{figure}

\clearpage  

\begin{figure}
\caption{Color-magnitude diagram for the YSO candidates in W3 Main. 
Class II candidates are indicated by stars, filled triangles represent Class I 
candidates, and the filled circles are red sources with H-K $>$ 2 having 
J counterparts. The solid curve denotes the loci of 10$^6$ yr old PMS stars, 
and the dotted curve for those of 0.3$\times10^{6}$ yr old ones, both derived 
from the model of Palla \& Stahler (1999). Masses range from 0.1 to 
4 M$_{\odot}$ from bottom to top, for both curves. The solid oblique reddening
line denotes position of PMS with 2 M$_{\odot}$ for 1 Myr and the dotted
oblique lines denote positions of PMS with 2 and
4 M$_{\odot}$ for 0.3 Myr, respectively, in this diagram. Most of the objects
well above the PMS tracks are luminous and massive ZAMS stars (see Table 1 and
Fig. 4).
\label{fig9}}
\end{figure}   

\begin{figure}
\caption{Enlarged view of the color image of selected areas (see Fig. 2 and
Appendix). a) a circular bracelike diffuse emission probably illuminated by
two bright stars of the same luminosity located toward east of this feature.
b) H$_2$ knot reminiscent of HH object which surrounds at least three YSOs.
c) an isolated red source at the center of the image, which is located
towards the south-west of IRS 4. The source is resolved into double stars
separated by $\sim$ 4\arcsec. d) dark filamentary lanes with irregular shapes.
An infrared source, marked by an arrow, 
with large color excesses is located inside one of them. 
\label{fig10}}
\end{figure}


\begin{thebibliography}{}
\bibitem[]{735}Bertout, C., Basri, G., \& Bouvier, J. 1988, \apj, 330, 350
\bibitem[]{736}Bessell, M. S., \& Brett, J. M. 1988, \pasp, 100, 1134
\bibitem[]{} Carpenter, J. M., Snell, R. L., Schloerb, F. P., et al. 1993,
\apj, 407, 657
\bibitem[]{737}Carpenter, J. M., Heyer, M. H., \& Snell, R. L. 2000, \apjs, 130, 381
\bibitem[]{738}Claussen, M. J., Gaume, R. A., Johnston, K. J., \& Wilson, T. L. 1994,
\apjl, 424, L41
\bibitem[]{741}Dickel, H. R., Dickel, J. R., Wilson, W. J., \& Werner, M. W. 1980,
\apj, 237, 711
\bibitem[]{743}Dyck, H. M., \& Simon, T. 1977, \apj, 211, 421
\bibitem[]{744}Elmegreen, B. G., \& Lada, C. J. 1977, \apj, 214, 725
\bibitem[]{745}Forster, J. R., Welch, W. J., \& Wright, M. C. H. 1977, \apjl, 215, L121
\bibitem[]{746}Gaume, R. A., \& Mutel, R. L. 1987, \apjs, 65, 193
\bibitem[]{747}Harris, S., \& Wynn-Williams, C. G. 1976, \mnras, 175, 649
\bibitem[]{748}Hofner, P., Delgado, H., Whitney, B., Churchwell, E., \&
Linz, H. 2002, \apjl, 579, L95
\bibitem[]{750}Imai, H., Kameya, O., Sasao, T., et al. 2000, \apj, 538, 751
\bibitem[]{751}Jiang, Z., Yao, Y., Yang, J., et al. 2002, \apj, 577, 245
\bibitem[]{752}Karr, J. L., \& Martin, P. G. 2003, \apj, 595, 900
\bibitem[]{}Koornneef, J. 1983, A\&A, 128, 84
\bibitem[]{753}Lada, E. A., Evans, N. J., Depoy, D. L., \& Gatley, I. 1991, 
\apj, 371, 171
\bibitem[]{754}Lada, C. J., \& Lada, E. A. 1991, ASP Conf. Ser. Vol. 13, p3
\bibitem[]{755}Lada, C. J., Young, E. T., \& Greene, T. P. 1993, \apj, 408, 471
\bibitem[]{758}Lada, C. J., \& Lada, E. A. 2003, \araa, 41, 57
\bibitem[]{756}Ladd, E. F., Deane, J. R., Sanders, D. B., \& 
Wynn-Williams, C. G. 1993, \apj, 419, 186
\bibitem[]{759}Megeath, S. T., Herter, T., Beichman, C., Gautier, N., 
Hester, J. J., Rayner, J., \& Shupe, D. 1996, \aap, 307, 775 
\bibitem[]{761}Meyer, M., Calvet, N., \& Hillenbrand, L. A. 1997, \aj, 114, 288
\bibitem[]{762}Nagashima, C., et al. 1999, Proc. Star Formation 1999, ed. 
T. Nakamoto (Nagano: Nobeyama Radio Obs.), 387
\bibitem[]{764}Nagashima, C., Dobbie, P. D., Nagayama, T., et al. 2003,
\mnras, 343, 1263
\bibitem[]{766}Nagayama, T., et al. 2003, Proc. SPIE, 4841, 459
\bibitem[]{767}O'Dell, C. R. 2001, \araa, 39, 99
\bibitem[]{768}Palla, F., \& Stahler, S. 1999, \apj, 525, 772
\bibitem[]{769}Persi, P., Roth, M., Tapia, M., et al. 1994, \aap, 282, 474
\bibitem[]{770}Persson, S. E., Murphy, D. C., Krzeminski, W., Roth, M., \& 
Rieke, M. J. 1998, AJ, 116, 2475
\bibitem[]{772}Richardson, K. J., Sandell, G., White, G. J., Duncan, W. D., \&
Krisciunas, K. 1989, \aap, 221, 95 
\bibitem[]{774}Rieke, G. H., \& Lebofsky, M. J. 1985, \apj, 288, 618
\bibitem[]{775}Roberts, D. A., Crutcher, R. M., \& Troland, T. H. 1997, \apj, 479, 318
\bibitem[]{777}Sugitani, K., Tamura, M., Nakajima, Y., et al. 2002, \apjl, 565, L25
\bibitem[]{}Stetson, P. B. 1987, \pasp, 99, 191
\bibitem[]{778}Tamura, M., Itoh, Y., Oasa, Y., \& Nakajima, T. 1998, Science,
282, 1095
\bibitem[]{780}Tapia, M., Persi, P., Bohigas, J., \& Ferrari-Toniolo, M. 1997, 
\aj, 113, 1769
\bibitem[]{782}Thronson, H. A. J, Lada, C. J., \& Hewagama, T. 1985,
\apj, 297, 662
\bibitem[]{784}Tieftrunk, A. R., Gaume, R. A., Claussen, M. J., Wilson, T. L., \&
Johnston, K. J. 1997, \aap, 318, 931 (TGC97)
\bibitem[]{}Tieftrunk, A. R., Megeath, S. T., Wilson, T. L., \& Rayner, J. T.
1998, \aap, 336, 991
\bibitem[]{}Wainscoat, R. J., Cohen, M., Volk, K., et al. 1992, \apjs, 83, 111
\bibitem[]{786}Wilson, T. L., Boboltz, D. A., Gaume, R. A., \& Megeath, S. T.
2003, \apj, 597, 434
\bibitem[]{788}Wynn-Williams, C. G. 1971, \mnras, 151, 397
\bibitem[]{789}Wynn-Williams, C. G., Werner, M. W., \& Wilson, W. J. 1974, \apj, 187, 41
\end{thebibliography}
\end{document}